\documentclass[aps,prl,twocolumn,showpacs,superscriptaddress]{revtex4-1}

\usepackage[utf8]{inputenc}
\usepackage[T1]{fontenc}
\usepackage{lmodern}
\usepackage{amsmath,amsfonts,amssymb}
\usepackage{graphicx}

\usepackage{hyperref}
\usepackage[usenames,dvipsnames]{xcolor}
\hypersetup{colorlinks=true, linkcolor=blue!50!black, urlcolor=blue!50!black, citecolor=blue!50!black}
\usepackage[all]{hypcap}

\newcommand{\img}{\mathsf{i}}
\newcommand\diff{\mathrm{d}}
\newcommand{\Ps}{\text{Ps}}
\newcommand{\subref}[2]{\hyperref[#1]{#2}}
\renewcommand{\vec}[1]{\mathbf{#1}}
\renewcommand{\phi}[0]{\varphi}

\usepackage[normalem]{ulem}

\begin{document}

\title{Tube Concept for Entangled Stiff Fibers Predicts Their Dynamics in Space and Time}


\author{Sebastian Leitmann}
\affiliation{Institut f\"ur Theoretische Physik, Universit\"at Innsbruck, Technikerstra{\ss}e~21A, A-6020 Innsbruck, Austria}
\author{Felix H\"ofling}
\affiliation{Fachbereich Mathematik und Informatik, Freie Universität Berlin, Arnimallee 6, 14195 Berlin, Germany}
\author{Thomas Franosch}
\affiliation{Institut f\"ur Theoretische Physik, Universit\"at Innsbruck, Technikerstra{\ss}e~21A, A-6020 Innsbruck, Austria}
\email[]{thomas.franosch@uibk.ac.at}

\date{\today}

\begin{abstract}
We study dynamically crowded solutions of stiff fibers deep in the semidilute regime, where the motion of a single
constituent becomes increasingly confined to a narrow tube.  The spatiotemporal dynamics for wave numbers resolving the
motion in the confining tube becomes accessible in Brownian dynamics simulations upon employing a geometry-adapted
neighbor list.  We demonstrate that in such crowded environments the intermediate scattering function, characterizing
the motion in space and time, can be predicted quantitatively by simulating a single freely diffusing phantom needle
only, yet with very unusual diffusion coefficients.
\end{abstract}


\maketitle



Long, stiff filaments are abundant in nature and nano\-technology and form entangled meshworks of remarkable mechanical
response and complex dynamic behavior~\cite{Bausch:NatPhys_2:2006}. Examples include biopolymers such as
filamentous actin (F-actin)~\cite{Liu:PRL_96:2006, Koenderink:PRL_96:2006, Wong:PRL_92:2004},
microtubuli~\cite{Lin:MA_40:2007}, \textit{fd} viruses~\cite{Lettinga:PRL_99:2007,Addas:PRE_70:2004}, and
xanthan~\cite{Koenderink:PRE_69:2004}, as well as more recently synthetically fabricated carbon nanotubes, cellulose
whiskers, and polymeric stiff rods~\cite{Cassagnau:Rheol_52:2013}.  In the \emph{semidilute} regime the excluded volume becomes
irrelevant,  such that the filaments in solution diffuse as
infinitely thin needles and their peculiar transport properties emerge due to topological constraints imposed by their neighboring
impenetrable filaments.
As a consequence, the motion of a single constituent in the highly entangled regime is
suppressed to a sliding back-and-forth movement in an effective tube composed of the surrounding filaments.  This \emph{tube
concept}, pioneered by Doi and Edwards~\cite{Doi:JP_36:1975, Doi:JCSFT_74:1978}, thus reduces the complex many-body
dynamics to an effective single-particle motion. Most prominently, it predicts a drastic suppression
of the rotational diffusion coefficient~\cite{Doi:JP_36:1975,Doi:JPSJ_53:1984,Tao:JCP_124:2006,Tse:JCP_139:2013}
as well as of the translational diffusion coefficient perpendicular to the orientation of the stiff filament~\cite{Teraoka:JCP_89:1988,Szamel:PRL_70:1993}.

Rod and needle systems have been studied in simulations and theory extensively~\cite{Loewen:PRE_50:1994, Tucker:JPCA_114:2010,Sussman:PRL_107:2011, Yamamoto:ACSML_4:2015}, differing in the underlying dynamics and aspect ratio of the
constituents. For Newtonian dynamics, the transport coefficients behave rather
differently~\cite{Frenkel:PRL_47:1981,Frenkel:MolPhys_49:1983,Otto:JCP_124:2006,Hoefling:PRL_101:2008,Tucker:JPCB_115:2011},
characterized by an increase in the center of mass diffusion coefficient, first observed by Frenkel and
Maguire~\cite{Frenkel:PRL_47:1981}.  A long-standing debate how a finite aspect ratio or a small flexibility of the
stiff Brownian rods affects the Doi-Edwards scaling~\cite{Doi:JPSJ_53:1984,Fixman:PRL_54:1985}, has been resolved only
recently~\cite{Cobb:JCP_123:2005,Tao:JCP_124:2006,Tse:JCP_139:2013,Zhao:Poly_54:2013}, pointing out, that the densities considered so far
just reach the onset of the asymptotic regime and corrections to scaling are relevant~\cite{Tse:JCP_139:2013}. Within
two-dimensional toy models~\cite{Moreno:EPL_67:2004, Hoefling:PRE_77:2008,Munk:EPL_85:2009}, higher densities can be achieved in the simulation and the Doi-Edwards picture has been
shown to remain valid.
Yet, extracting spatiotemporal information, in principle accessible in
light-scattering experiments~\cite{Doi:JCSFT_74:1978,Aragon:JCP_82:1985} or simulations, has not been an easy task and
the fundamental implications for the motion in the tube have remained unexplored.

Here, we show that the tube concept can be elaborated to quantitatively predict the intermediate scattering function of highly
entangled suspensions of stiff filaments at all time and length scales.
We also compare to a needle Lorentz system, where a single tracer needle explores a \emph{quenched} disordered array of needles, i.e.
the proverbial needle in a haystack, and demonstrate that the dynamic rearrangement of the surrounding needles induces
no qualitative changes. We corroborate the scaling behavior of the transport coefficients and show that the
spatiotemporal dynamics
can be rationalized in terms of a single phantom needle where transport is dominated by translation-rotation coupling.
For very high entanglement analytic progress is made and we show that the intermediate scattering function can be evaluated in closed form for
long times.

\textit{Transport coefficients.---}
In the semidilute regime the stiff filaments can be considered as infinitely thin needles characterized solely by their length $L$,
leading to the dimensionless number density $n^* = nL^3$ as only structural control parameter.
We investigate the dynamics of a suspension of needles relying on stochastic simulations where the needles undergo
diffusion with short-time rotational diffusion coefficient
$D_\text{rot}^0$ and short-time translational diffusion coefficient for parallel $D_\parallel^0$ and perpendicular
$D_\perp^0$ motion with respect to the needle axis.
The hard-core interaction between the needles is handled following a pseudo-Brownian
scheme~\cite{Scala:JCP_126:2007, Frenkel:MolPhys_49:1983} and a geometry-adapted neighbor list to speed
up the collision detection by up to two orders of magnitude (see Supplemental Material).

\begin{figure}[t]
\includegraphics[scale=0.5]{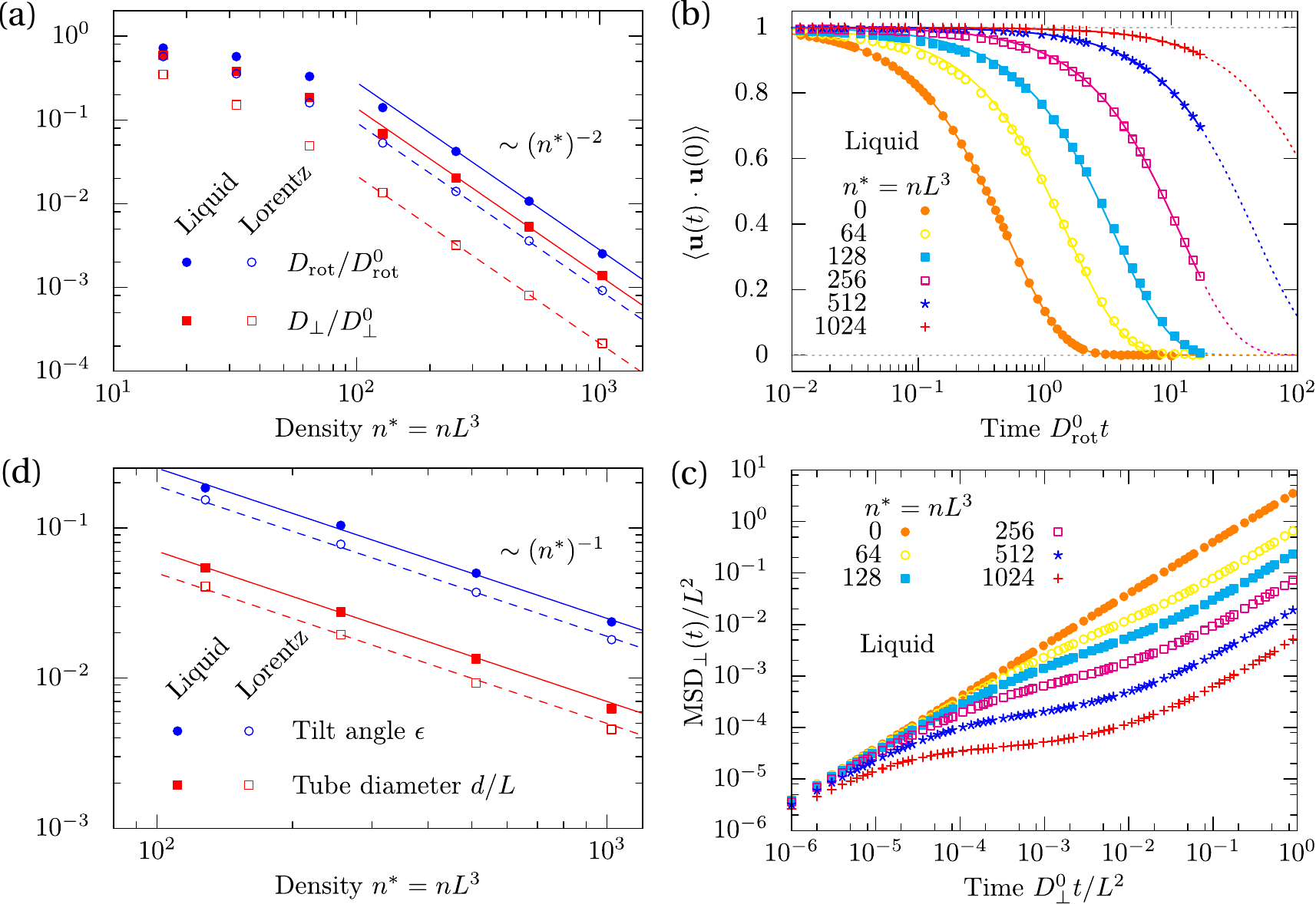}
\caption{ \label{fig:1}
Translational and rotational transport for needle liquids and needle Lorentz systems. (a)
Suppression of the rotational $D_\text{rot}$ and translational $D_\perp$ diffusion coefficient as a function of the needle density.
(b) Correlation function $\langle \vec{u}(t)\cdot\vec{u}(0)\rangle$ for the orientation
$\vec{u}(t)$ of the needle axis.
(c) Mean-square displacement $\text{MSD}_\perp(t)$ measured in the body-fixed frame perpendicular to the needle axis.
(d) Dimensionless tube diameter $d/L$ and tilt angle $\epsilon$ in the scaling regime. }
\end{figure}

High entanglement occurs deep in the
semidilute regime $n^* \gg 1$ and the motion of a single needle becomes strongly confined to a narrow tube and both the
rotational diffusion coefficient $D_\text{rot}$~\cite{Doi:JP_36:1975,Doi:JPSJ_53:1984,Tao:JCP_124:2006,Tse:JCP_139:2013}
as well as the translational diffusion coefficient perpendicular to the orientation of the rod $D_\perp$ are anticipated
to scale as $\sim (n^*)^{-2}$~\cite{Teraoka:JCP_89:1988,Szamel:PRL_70:1993}.
The rotational diffusion coefficient $D_\text{rot}$ extracted from simulations is shown in Fig.~\subref{fig:1}{1(a)} deep in the semidilute
regime.  The transport coefficient is suppressed by a factor of $10$ up to densities of $n^* \approx 100$, the highest
density which has been considered before~\cite{Doi:JPSJ_53:1984, Tse:JCP_139:2013}. Our data extend to densities of
$n^* \approx 1000$ such that the rotational diffusion is suppressed by another factor of $100$. The Doi-Edwards
prediction $\sim (n^*)^{-2}$ is nicely followed for the high densities $n^* \gtrsim 100$. The data for the needle
Lorentz system are qualitatively similar, in the highly entangled regime the scaling prediction is nicely corroborated,
however, the static environment reduces the rotational diffusion by another factor of $\approx 3$.

The time-dependent correlation function $\langle \vec{u}(t)\cdot\vec{u}(0)\rangle$ for the unit vector of orientation
$\vec{u}(t)$ of the needle slows down drastically as the density is increased~[Fig.~\subref{fig:1}{1(b)}]. The shape of the
longtime relaxation is well represented by an exponential, $\exp(-2D_\text{rot} t)$, characterized by the rotational
diffusion coefficient $D_\text{rot}$.  We have
also measured higher-order orientational correlation functions $\langle \text{P}_\ell(\vec{u}(t)\cdot\vec{u}(0))
\rangle$, $\ell = 2, 3$, where $\text{P}_\ell(\cdot)$ denotes the Legendre polynomials, and found that they also decay
as $\exp[-\ell(\ell+1)D_\text{rot}t]$ with the same transport coefficient. Therefore, we conclude that at long times the
pure orientational motion corresponds to simple diffusion on a sphere.  At
intermediate times $d^2/D_\perp^0 \ll t \ll (D_\text{rot})^{-1}$ when the needle is confined to its initial tube with
diameter $d$, the orientational correlation function displays a plateau $\cos(\epsilon)$ close to unity, which we use to
extract the tilt angle $\epsilon$ displayed in Fig.~\subref{fig:1}{1(d)}.

Furthermore, we have measured the mean-square displacements in the body-fixed frame. Since a collision does not affect the dynamics parallel to the needle,
the corresponding mean-square displacement $\text{MSD}_\parallel(t) = 2D_\parallel^0 t$ is trivial, in particular, the parallel diffusion coefficient
is density independent, $D_\parallel = D_\parallel^0$.
In contrast, the mean-square displacement $\text{MSD}_\perp(t)$ in the body-fixed frame perpendicular to the needle axis
becomes diffusive only at long times [Fig.~\subref{fig:1}{1(c)}],
from which we obtain the perpendicular diffusion
coefficient $D_\perp$. The diffusion coefficient $D_\perp$ is suppressed by up to three orders of magnitude as can be inferred from
Fig.~\subref{fig:1}{1(a)}, following the scaling prediction $D_\perp \sim (n^*)^{-2}$~\cite{Teraoka:JCP_89:1988, Szamel:PRL_70:1993}. For high needle
densities, the data for the mean-square displacement $\text{MSD}_\perp(t)$ [Fig.~\subref{fig:1}{1(c)}] display an extended plateau
$\text{MSD}_\perp \approx d^2$ which we use to measure the tube diameter $d$ shown in Fig.~\subref{fig:1}{1(d)}, scaling as $(n^*)^{-1}$
with needle density. For the needle Lorentz system, the data for the rotational diffusion and perpendicular translation
are qualitatively identical to the needle liquid.

\begin{figure*}[t]
\includegraphics[scale=0.68]{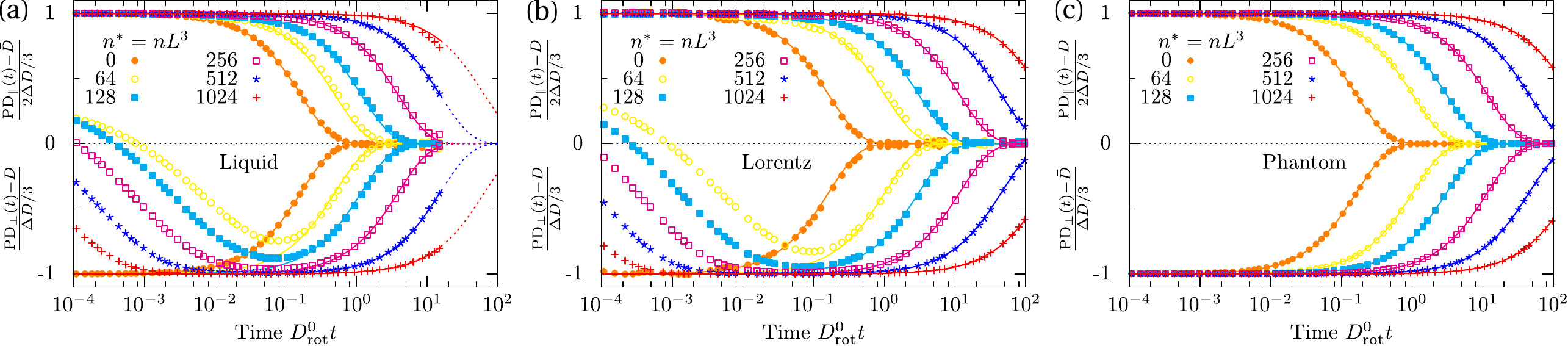}
\caption{ \label{fig:2}
Translation-rotation coupling measured in terms of the deviation of the projected time-dependent diffusion coefficients
$\text{PD}_\parallel(t)$ and $\text{PD}_\perp(t)$ from their longtime limits $\bar{D} = (D_\parallel + 2D_\perp)/3$
normalized by the diffusional anisotropy $\Delta D = D_\parallel - D_\perp$. Lines represent analytic results $\pm
\exp(-6D_\text{rot}t)$ for translation-rotation coupling from anisotropic diffusion.  Data are shown for (a) needle liquid
and (b) needle Lorentz systems, as well as (c) a freely diffusing needle (phantom needle), using the measured transport coefficients from the
simulation of a needle Lorentz system.
}
\end{figure*}

\textit{Translation-rotation coupling.---} As a second step we investigate the relaxation of the translation-rotation coupling in terms of the tube model.
The translational dynamics is anticipated to be transiently highly anisotropic, since diffusion occurs essentially
along the needle axis. To quantify the translation-rotation coupling we measure the displacements of the needle center
$\Delta \vec{r}(t) =
\vec{r}(t) - \vec{r}(0)$ projected along or perpendicular to its initial axis $\vec{u}(0)$ and average its square. We denote the
corresponding time derivatives
\begin{align}
\text{PD}_\parallel(t) &= \frac{1}{2}\frac{\diff}{\diff t} \langle [\vec{u}(0)\cdot\Delta\vec{r}(t)]^2\rangle \ , \\
\text{PD}_\perp(t) &= \frac{1}{4}\frac{\diff}{\diff t} \langle [\vec{u}(0)\times\Delta\vec{r}(t)]^2 \rangle\ ,
\end{align}
as projected time-dependent diffusion coefficients. For short times, $t \lesssim 10^{-2}(D_\text{rot})^{-1}$ the projected diffusion
coefficients display pure parallel diffusion $\text{PD}_\parallel(t) \approx D_\parallel$, since the needle axis remains aligned with its
initial orientation [see Fig.~\subref{fig:2}{2(a)}]. In contrast, for long times  $t \gtrsim (D_\text{rot})^{-1}$ the initial orientation is forgotten and the motion becomes
isotropic diffusion $\text{PD}_\parallel(t\to\infty) = \bar{D} := (D_\parallel + 2D_\perp)/3$. For the projected perpendicular diffusion, in addition to the
short-time plateau $\text{PD}_\perp(t) \approx D_\perp^0$ for times $t \lesssim d^2/D_\perp^0$, a new plateau $\text{PD}_\perp(t) \approx D_\perp$
emerges for intermediate times $d^2/D_\perp^0 \ll t \ll (D_\text{rot})^{-1}$, reflecting the strong suppression of the perpendicular motion
by the confining tube. The tube constraint is relaxed only at much longer times $t \gtrsim (D_\text{rot})^{-1}$ and the projected perpendicular
dynamics reaches isotropic diffusion $\text{PD}_\perp(t\to\infty) = \text{PD}_\parallel(t\to\infty) = \bar{D}$. The
needle Lorentz system shown in Fig.~\subref{fig:2}{2(b)} displays the same
phenomenology. We compare the translation-rotation coupling induced by the tube to a \emph{phantom} needle diffusing freely in three-dimensional space,
relying on the transport coefficients measured at long times. Thus, the phantom needle undergoes anisotropic diffusion in a rather peculiar effective medium such that
the perpendicular diffusion and the rotational motion are suppressed by orders of magnitude with respect to the bare parallel diffusion $D_\parallel^0$.
The corresponding projected time-dependent diffusion coefficients, displayed in Fig.~\subref{fig:2}{2(c)}, reproduce quantitatively the translation-rotation coupling both for a needle liquid as well as for the needle Lorentz system, provided the intermediate plateau becomes manifest. Hence, we conclude that also the coupling between
the orientational and translational motion is well accounted for in terms of effective anisotropic diffusion. The coupling for the phantom needle can
be worked out analytically, for example, by extending the methods developed by Han \emph{et~al.}~\cite{Han:Science_314:2006} for the free two-dimensional anisotropic diffusion of ellipsoidal particles. For the three-dimensional case, we obtain
\begin{align}
\text{PD}_\parallel(t) = \bar{D} + \frac{2}{3}\, \Delta D\, \exp(-6 D_\text{rot} t), \\
\text{PD}_\perp(t) = \bar{D}  - \frac{1}{3}\, \Delta D\, \exp(-6 D_\text{rot} t),
\end{align}
where $\Delta D = D_\parallel - D_\perp$ characterizes the diffusional anisotropy.
The theoretical curves are included in Fig.~\ref{fig:2} and are in excellent agreement with the simulation data.

\begin{figure*}[t]
\includegraphics[scale=0.7]{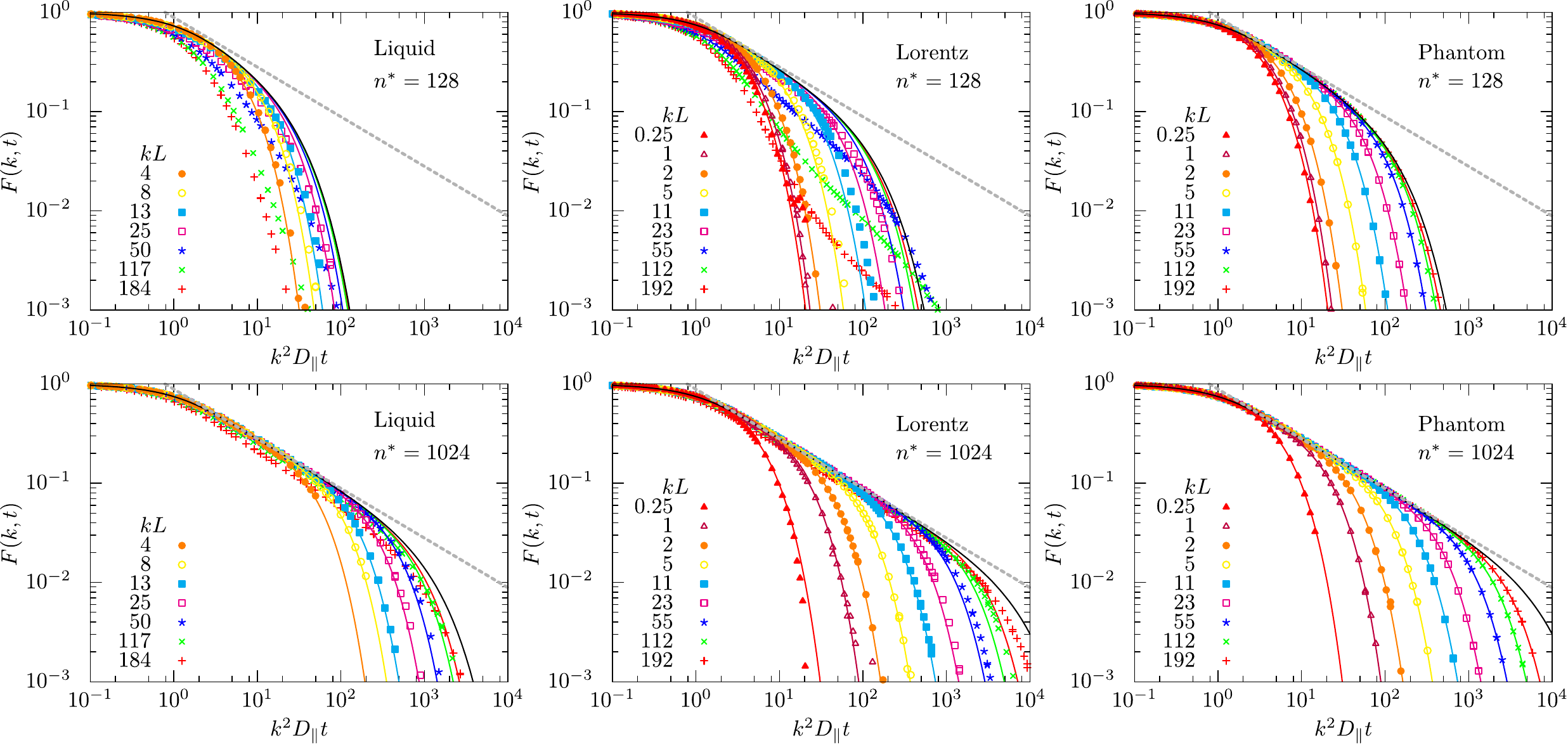}
\caption{ \label{fig:3}
Intermediate scattering function for a broad range of wave numbers $k = |\vec{k}|$. The solid lines correspond to the
analytic predictions of the tube model~[Eq.~\eqref{eq:intermediate_scattering_sum_spheroidal}] relying on the extracted
transport coefficients. The phantom needle is for the measured diffusion coefficients of the needle Lorentz system. The
bottom panels show data for the high entanglement. As guide to the eye, the straight dashed lines indicate power law
decays $\sim t^{-1/2}$ emerging from the sliding motion in the tube. 
}
\end{figure*}

\textit{Intermediate scattering function.---}
The success of the tube model for the scaling behavior of the transport coefficients and the translation-rotation
coupling encourages us to explore the full ramifications of the tube model for the experimentally accessible
intermediate scattering function $F(k,t)$, providing temporal information on the needle dynamics on a length scale of
$2\pi/k$.  For simplicity we discuss only the motion of the needle center \begin{align} F(k,t) = \langle \exp[-\img
\vec{k}\cdot\Delta\vec{r}(t)]\rangle, \end{align} for different wave numbers $k = |\vec{k}|$. The data
(Fig.~\ref{fig:3}) display a statistical accuracy down to $10^{-3}$ and extend over up to five significant decades in
time. The data for different wave numbers collapse for short diffusion times $k^2 D_\parallel t$, since the needle
merely diffuses along its initial tube. For increasing wave number the data approach an emergent master curve with a
power-law decay $\sim t^{-1/2}$ for long times. In the highly entangled regime the data for the needle Lorentz system
are indistinguishable from the phantom needle at all wave numbers shown.

Since the intermediate scattering function for the needle liquid and the needle Lorentz system is in perfect agreement with the simulation
for the phantom needle, we also elaborate explicit analytic expressions for $F(k,t)$ within the tube model.
The stochastic dynamics of the phantom needle is completely encoded in the conditional probability $G(\vec{r},
\vec{u}, t| \vec{u}_0)$ for the needle to displace by $\vec{r}$ and change from an initial orientation $\vec{u}_0$ to
$\vec{u}$ in lag time $t$.
The time evolution of the conditional probability is then governed by a Smoluchowski-Perrin equation~\cite{Doi:JCSFT_74:1978,Doi:Oxford:1999} and its
spatial Fourier transform $G_\vec{k}(\vec{u}, t | \vec{u}_0) = \int\diff^3r\  e^{-\img
\vec{k}\cdot\vec{r}} G(\vec{r}, \vec{u}, t| \vec{u}_0)$ evolves according to
\begin{align} \label{eq:smoluchowski-perrin_k}
  \partial_t G_\vec{k} = D_\text{rot}\mathcal{R}\cdot(\mathcal{R}G_\vec{k}) -  \{k^2 D_\parallel - \Delta D [k^2 -
(\vec{k}\cdot\vec{u})^2]\}G_\vec{k},
\end{align}
with $k = |\vec{k}|$ and the rotational operator $\mathcal{R} = \vec{u} \times \partial_\vec{u}$.
The intermediate scattering function $F(k,t)$ is then obtained by averaging over all initial orientations $\vec{u}_0$
and integrating over all final orientations $\vec{u}$:
\begin{align} \label{eq:intermediate_scattering_function}
  F(k, t) = \int\!\diff^2 u \int\!\frac{\diff^2 u_0}{4\pi}\ G_\vec{k}(\vec{u}, t | \vec{u}_0) .
\end{align}

Here, we rely on a numerical evaluation of $F(k,t)$ in terms of eigenfunctions of the operator on the right-hand side
of Eq.~\eqref{eq:smoluchowski-perrin_k} which turn out to be spheroidal wave functions~\cite{Aragon:JCP_82:1985} (see
Supplemental Material). 
The integral in Eq.~\eqref{eq:intermediate_scattering_function} can be performed for each term in the expansion of
$G_\vec{k}$ and reduces to an evaluation of the projection of the zeroth order spheroidal wave functions of degree $n$
onto the zeroth degree Legendre polynomial, $a_{n,-n/2}^0$:
\begin{align} \label{eq:intermediate_scattering_sum_spheroidal}
   F(k, t) = e^{-k^2 D_\parallel t}\sum_{\substack{n = 0\\ n \text{ even}}}^\infty (2n + 1)
\bigl(a_{n,-n/2}^0\bigr)^2 e^{-\lambda_n^0 D_\text{rot} t},
\end{align}
with spheroidal eigenvalue $\lambda_n^0$. In essence, the intermediate scattering function $F(k,t)$ is a weighted average
of relaxing exponentials only, i.e. a completely monotone function~\cite{Feller:Wiley:1970}. In particular, the longtime relaxation is purely exponential with a rate constant
provided by the lowest eigenvalue of Eq.~\eqref{eq:smoluchowski-perrin_k}.

In the highly entangled regime $D_\text{rot}$ becomes small and for wave numbers fulfilling $\gamma^2 := k^2 \Delta D/D_\text{rot} \gg 1$ the orientational motion in
Eq.~\eqref{eq:smoluchowski-perrin_k} is dominated by translational diffusion. Then, one can show that the spheroidal eigenvalues approach
$\lambda_n^0 = -\gamma^2 + (2n+1)\gamma + \mathcal{O}(1)$ and the spheroidal wave functions are approximated by the eigenfunctions
of the harmonic oscillator for fixed $n \ll \gamma$~\cite{Meixner:Springer:1954}. Then, for times $D_\text{rot} \gamma^2 t \gtrsim 1$ only terms where the approximation is valid contribute
to the sum of Eq.~\eqref{eq:intermediate_scattering_sum_spheroidal} such that the sum can be evaluated in closed form
(see Supplemental Material) to
\begin{align} \label{eq:intermediate_scattering_simplified}
F(k,t) = e^{-k^2 D_\perp t} \sqrt{\frac{\pi}{\gamma}} \frac{1}{\sqrt{2\sinh(2 D_\text{rot}\gamma t)}} .
\end{align}

We have checked that Eq.~\eqref{eq:intermediate_scattering_simplified} is an accurate representation of the intermediate
scattering function for $\gamma^2 \gtrsim 180\ (n^* = 1024)$, yet the full solution faithfully represents
the simulation data already at smaller values $\gamma^2 \lesssim 50\ (n^* = 128)$. For times $D_\text{rot}\gamma t \lesssim 1$, our analytic expression simplifies to
\begin{align} \label{eq:intermediate_asymptotic}
F(k, t) = \frac{e^{-k^2 D_\perp t}}{\sqrt{4 k^2 \Delta D t/\pi}} ,
\end{align}
which rationalizes the power-law tail observed in the simulation data (Fig.~\ref{fig:3})
as well as the data collapse for different wave numbers and times  $k^2 D_\perp t \lesssim 1$.

\textit{Summary and conclusion.---}
The spatiotemporal dynamics of a needle in a highly entangled suspension has been characterized in terms of
the transport coefficients, the time-dependent translation-rotation coupling, and the intermediate scattering function
for a broad range of wave numbers and times. Relying on a geometry-adapted neighbor list, collisions can be handled
efficiently for dynamical crowded systems. In this regime, our simulation data collapse to the dynamics of a single
phantom needle with renormalized transport coefficients, thus confirming for the first time the relevance of the tube
model for the spatiotemporal dynamics.

Our data demonstrate that the notion of the tube concept entails predictive power for the intermediate scattering
function for both needle Lorentz systems and needle liquids.  In the latter case, the mobile constraints cause a
dilation of the effective tube width and the tube dynamically reorganizes by \emph{constraint release} 
processes~\cite{Doi:Oxford:1999}, which occur on the same time scale $L^2/D_\parallel$ as the polymer disengages from
its initial tube. Comparing our simulation results for the two cases confirms that constraint release only shifts the prefactors of the
scaling law $D_\text{rot} \sim (n^*)^{-2}$, thus speeding up the dynamics. Hence, the dynamics in the needle liquid is identical
to that of the needle Lorentz albeit at a lower density. Yet, the reduction of the many-body problem to the single
freely diffusing phantom needle is robust, as suggested by Doi and Edwards~\cite{Doi:Oxford:1999}.

We have provided explicit analytic expressions in terms of spheroidal wave functions and their asymptotic harmonic
oscillator behavior including the long algebraic tail, a fingerprint of the sliding motion within the tube.
Interestingly, the tube model becomes quantitative already at densities and wave numbers where the tail is not yet
manifest. Correspondingly, to account for the dynamics for all times and wave numbers, one has to resort to the full
solution.

In principle, in scattering experiments not only the center of the needle is relevant but the entire rod contributes to
the scattering signal. In essence, a form factor $f_\vec{k}(\vec{u}) =
\sin(\vec{k}\cdot\vec{u}L/2)/(\vec{k}\cdot\vec{u}L/2)$ has to be included ~\cite{Doi:Oxford:1999} which considerably
complicates the analytic solution~\cite{Aragon:JCP_82:1985}.  To obtain closed expressions for the modified intermediate
scattering functions, Doi and Edwards~\cite{Doi:JCSFT_74:1978} also rely on the analogy to the harmonic oscillator but
use an additional technical approximation which is circumvented in our analysis (see Supplemental Material). In
simulations, the modified intermediate scattering function can be evaluated easily in terms of the phantom needle, yet,
the overall picture remains unchanged since the dynamics is dominated by the back and forth motion in the narrow
confining tube.

In the context of semiflexible polymers in the highly entangled regime, the tube concept has also proved itself to be
the key insight~\cite{Odijk:MA_16:1983, Semenov:JCSFT_82:1986}, although its ramifications have been tested mostly for
static properties~\cite{Romanowska:EPL_86:2009, Morse:PRE_63:2001}.  It is anticipated to still capture the dynamics at
coarse-grained time and length scales~\cite{Ramanathan:PRE_76:2007,Nam:JCP_133:2010,Keshavarz:ACSN_10:2016}, yet the
tube displays additional tube-width fluctuations~\cite{Glaser:PRL_105:2010,Wang:PRL_104:2010} and a more complex tube
renewal.  Therefore, in the semiflexible case constraint release and tube renewal might admit a faster terminal
relaxation such that the algebraic tail in the intermediate scattering function may be difficult to observe. One may
speculate that the most important change is again a modified scaling law for the transport coefficients, while the
semiflexible polymer still diffuses in and out of its tube along its own
backbone~\cite{Kaes:Nature_368:1994,Fakhri:Science_330:2010}.

The one-dimensional sliding motion should, in fact, also be the most important ingredient for the dynamics of
suspensions of nanofibers with finite aspect ratio and flexibility, and we anticipate that these systems also become
accessible in simulations relying on our novel algorithm. With the rapid technological advancement to fabricate stiff
nanorods and nanofibers~\cite{Cassagnau:Rheol_52:2013,Kasimov:PRE_93:2016}, the semidilute regime of highly entangled suspensions has become
finally into experimental reach.

\begin{acknowledgments}
This work was supported by the Austrian Ministry of Science BMWF as part of the UniInfrastrukturprogramm of the Focal
Point Scientific Computing at the University of Innsbruck. We acknowledge financial support by the Deutsche
Forschungsgemeinschaft (DFG) Contract No. FR1418/5-1.
\end{acknowledgments}


%

\onecolumngrid
\clearpage

\section{Supplemental Material}

\subsection{Stochastic simulation}

We consider the diffusion of an infinitely thin needle of length $L$ with the short-time rotational 
diffusion coefficient $D_\text{rot}^0$ and the short-time translational diffusion coefficient for parallel
$D_\parallel^0$ and perpendicular $D_\perp^0$ motion. At every time $t$, the state of the needle is completely described
by the position of the center
of the needle $\vec{r}(t)$ and the unit vector of the orientation $\vec{u}(t)$.  The starting point for the
stochastic simulation are the following Langevin equations in the It\={o} interpretation for the change in orientation $\diff\vec{u}$ and
position $\diff\vec{r}$~\cite{Chirikjian:Birkhaeuser:2009, Oksendal:Springer:2010}:
\begin{align}   
\begin{split} \label{eq:langevin_equation}
  \diff\vec{u} &= -2 D_\text{rot}^0 \vec{u} \diff t - \sqrt{2 D_\text{rot}^0} \vec{u} \times \boldsymbol\xi \diff t, \\
  \diff\vec{r} &= \big[ \sqrt{2 D_\parallel^0} \vec{uu} + \sqrt{2 D_\perp^0} (1 - \vec{uu})
\big]\boldsymbol\eta \diff t .
\end{split}
\end{align}
The independent Gaussian white noise processes $\boldsymbol\xi$ and
$\boldsymbol\eta$ with zero mean and covariance $\langle \xi_i(t) \xi_j(t') \rangle =\langle
\eta_i(t)\eta_j(t')\rangle = \delta_{ij}\delta(t - t')$ 
determine the stochastic dynamics. The dyadic product is denoted by $\vec{uu}$.

We implement the Langevin equations using a discrete fixed Brownian time step $\tau_\text{B}$ and assign to the needle
the random pseudo-angular velocity $\boldsymbol\omega$ and pseudo-translational velocity $\vec{v}$ according to 
\begin{align}
\begin{split} \label{eq:pseudo-velocities}
\boldsymbol\omega &= \sqrt{\frac{2 D_\text{rot}^0}{\tau_\text{B}}}(1 - \vec{uu})\boldsymbol{\mathcal{N}}_\xi ,\\
\vec{v} &= \big[ \sqrt{\frac{2D_\parallel^0}{\tau_\text{B}}} \vec{uu} + \sqrt{\frac{2D_\perp^0}{\tau_\text{B}}} (1 - \vec{uu}) \big]\boldsymbol{\mathcal{N}}_\eta.
\end{split}
\end{align}
The Gaussian
white noise process are expressed via the independent and normal distributed
random variables $\boldsymbol{\mathcal{N}}_\xi$ and
$\boldsymbol{\mathcal{N}}_\eta$ with zero mean and unit variance. 

During a Brownian time step $\Delta t \in [0,\tau_\text{B}]$, the needle evolves ballistically according to the propagation rules
\begin{align}
\begin{split} \label{eq:propagation_rules}
  \vec{u}(t + \Delta t) &= \vec{u}(t)\cos(|\boldsymbol\omega| \Delta t) +
\biggl(\frac{\boldsymbol\omega}{|\boldsymbol\omega|}\times \vec{u}\biggr)\sin(|\boldsymbol\omega|
\Delta t),\\
  \vec{r}(t + \Delta t) &= \vec{r}(t) + \vec{v}\Delta t.
\end{split}
\end{align}

We simulate both needle liquids as well as needle Lorentz systems. In the former case, we move every
needle subsequently and fix all other needles, whereas in Lorentz systems only a single needle diffuses 
in a frozen array of other needles. 
For needle liquids we use a Brownian time-scale of $\tau_\text{B} = 10^{-6}L^2/D_\perp^0$ and observe the
dynamics over six decades in time, whereas for needle Lorentz systems we use $\tau_\text{B} = 10^{-8}L^2/D_\perp^0$ and generate
trajectories over ten decades in time. 

The hard-core interaction of the moving needle with the other ones is handled following a pseudo-Brownian
scheme~\cite{Scala:JCP_126:2007, Frenkel:MolPhys_49:1983}.
Here we assume, that during a Brownian time step the needle propagates ballistically according to the propagation
rules~(\ref{eq:propagation_rules}) in
continuous time $t\in [0,\tau_\text{B}]$.
Collisions between two needles are only possible if the distance of the intersection point of both needle centers is
smaller than half the needle length. The number of possible candidates can be greatly reduced by keeping an efficient
neighbor list adapted to the needle problem. It turns out that considering a surrounding cylinder is way more
advantageous than a sphere conventionally used for spherical particles. Geometry-adapted neighbor lists have been
considered before for ellipsoidal particles~\cite{Donev:JCompPhys:2005}, however its full advantage becomes manifest for very high aspect ratios, in
particular needles leading to a speedup of up to two orders of magnitude for the highest densities considered. The
efficiency is determined by the diameter of the cylinder which can be decreased upon considering smaller Brownian
time-scales $\tau_\text{B}$. 

The fixed collision 
candidate needle with orientation $\vec{u}_\text{c}$ intersects the rotational plane of the moving needle
perpendicular to $\boldsymbol\omega$ for times $\Delta t$ which fulfill 
\begin{align}
|\boldsymbol\omega\cdot \Delta\vec{r}_\text{c}(t + \Delta t)| \leq \frac{L}{2}|\boldsymbol\omega\cdot \vec{u}_\text{c}|, 
\end{align}
where the distance between the needle centers at time $t + \Delta t$ is denoted by $\Delta\vec{r}_\text{c}(t + \Delta t)$.
In addition to that, the intersection point of the obstacle needle with the rotational plane has to traverse the
rotational disk of radius $L/2$ of the moving needle. This is described as solution of the quadratic equation
\begin{align}
[\Delta \vec{r}_\text{pl}(t)]^2 = \Bigl(\frac{L}{2}\Bigr)^2,
\end{align}
where we denote the distance of the center of the moving needle with the intersection point in the rotational plane at
time $t$ as $\Delta \vec{r}_\text{pl}(t)$.
As a result the time interval for a possible collision reduces to the interval $[\tau_l, \tau_u] \subset [0,\tau_\text{B}]$.
If the new interval is nonempty, collision times $\Delta t\in [\tau_l, \tau_u]$ can be determined explicitly as solutions of the equation 
\begin{align}
     \boldsymbol \omega \cdot\bigl[\Delta\vec{r}_\text{pl}(t + \Delta t)\times\vec{u}(t)\bigr]\cos(\omega \Delta t) 
         + \omega\bigl[\Delta\vec{r}_\text{pl}(t + \Delta t)\cdot\vec{u}(t)\bigr]\sin(\omega \Delta t) = 0,
\end{align}
via an interval Newton method~\cite{Hansen:Dekker:2004}. Moreover, a correctly
rounded math library has to be used for the transcendental functions in order not to miss collisions. 

Upon colliding, the new velocities for translation and rotation are determined by energy, momentum and angular momentum
conservation.  We assume that the center of mass coincides with the geometric center of the needle. 
For smooth needles~\cite{Frenkel:MolPhys_49:1983, Otto:JCP_124:2006}, the momentum transfer onto the
moving needle $\Delta \vec{p} = \Delta p \vec{e}_m$ is directed perpendicular to both needle orientation and has a
magnitude of 
\begin{align} 
\Delta p = -2\frac{\vec{v}\cdot\vec{e}_m +
\boldsymbol\omega\cdot(\vec{r}_\text{coll}\times\vec{e}_m)}{1+\frac{m}{I}(\vec{r}_\text{coll}\times\vec{e}_m)^2}, 
\end{align}
where the distance between the point of contact and the center of the moving needle $\vec{r}_\text{coll}$ has been
introduced. In particular, the pseudo-velocity parallel to the needle does not change. 
The ratio of the mass $m$ to the moment of inertia $I$ remains a free parameter so far. We define a pseudo-temperature for the
rotational degrees of freedom by $I \langle \boldsymbol\omega^2\rangle/2 = 2 k_\text{B} T_\text{rot}/2$, and similarly
for the perpendicular translational degrees of freedom $m \langle \vec{v}_\perp^2\rangle/2 = 2 k_\text{B}T_\perp/2$. 
From the equation for the pseudo-velocities~(Eq. \ref{eq:pseudo-velocities}) we determine the averages 
$\langle \boldsymbol\omega^2\rangle = 4 D_\text{rot}^0/\tau_\text{B}$ and $\langle
\vec{v}_\perp^2\rangle = 4 D_\perp^0/\tau_\text{B}$. In order to avoid an average flow of energy between the 
rotational and the perpendicular translational degrees for freedom we impose $T_\text{rot} = T_\perp$ which yields
\begin{align}
\frac{m}{I} = \frac{D_\text{rot}^0}{D_\perp^0}.
\end{align}

In general, the transport coefficients of the needle are not independent and their relation is established by a hydrodynamic
calculation. For the simulation, we use the transport coefficients of a slender rod characterized by 
$D_\text{rot}^0 = 12D_\perp^0/L^2$ as well as $D_\parallel^0 = 2D_\perp^0$~\cite{Doi:Oxford:1999}.

\subsection{Intermediate scattering function of the phantom needle}

The stochastic dynamics of the phantom needle is translationally invariant in time and space and described by the 
the propagator $G(\vec{r}, \vec{u}, t| \vec{u}_0)$ which represents the conditional probability for a displacement $\vec{r}$
and change of orientation from $\vec{u}_0$ to $\vec{u}$ in lag time $t$. 
The time evolution of the propagator $G \equiv G(\vec{r},\vec{u},t|\vec{u}_0)$ with initial condition 
$G(\vec{r},\vec{u},t = 0|\vec{u}_0) = \delta(\vec{r})\delta(\vec{u},\vec{u}_0)$ 
is determined by the Smoluchowski equation
\begin{align} \label{eq:smoluchowski_anisotropic_dpara_dd}
  \partial_t G = D_\text{rot}\mathcal{R}\cdot(\mathcal{R}G) + \partial_\vec{r}\cdot[D_\parallel(\partial_\vec{r} G)
- \Delta D (1 - \vec{uu})(\partial_\vec{r} G)], 
\end{align}
with the rotational operator $\mathcal{R} = \vec{u} \times \partial_\vec{u}$~\cite{Berne:Dover:2000, Doi:Oxford:1999}.

The main quantity of interest for the dynamics is given by the self-intermediate
scattering function
\begin{align}
  F(\vec{k}, t) = \big\langle e^{-\img \vec{k}\cdot\Delta\vec{r}(t)} \big\rangle = \int\!\diff^2 u
\int\!\frac{\diff^2 u_0}{4\pi} G_\vec{k}(\vec{r}, \vec{u}, t | \vec{u}_0) 
\end{align}
as an average over all initial orientations $\vec{u}_0$ and integration over all final orientations $\vec{u}$ of the spatial Fourier
transform of the propagator $G(\vec{r},\vec{u},t|\vec{u}_0)$:
\begin{align}
  G_\vec{k}(\vec{u}, t | \vec{u}_0) = \int\!\diff^3r\  e^{-\img \vec{k}\cdot\vec{r}} G(\vec{r}, \vec{u}, t|
\vec{u}_0) .
\end{align}
The Smoluchowski equation for the spatial Fourier transform assumes the form
\begin{align}
  \partial_t G_\vec{k} = D_\text{rot}\mathcal{R}\cdot(\mathcal{R}G_\vec{k}) - \{k^2 D_\parallel - \Delta D [k^2 -
(\vec{k}\cdot\vec{u})^2]\}G_\vec{k} . 
\end{align}
We choose a coordinate system such that the wave vector is along the $z$-direction, $\vec{k} = |\vec{k}|\vec{e}_z$.
Then, the Smoluchowski-Perrin equation in spherical coordinates reads
\begin{align} \label{eq:smoluchowski-perrin_z_phi}
 \partial_t G_\vec{k}&= D_\text{rot}\{\partial_z[(1-z^2)\partial_z G_\vec{k}] + (1 - z^2)^{-1}\partial_\phi^2G_\vec{k}\} - k^2 [D_\parallel - \Delta D (1 - z^2)]G_\vec{k}\ ,
\end{align}
where the relation $z = \vec{k}\cdot\vec{u}/k = \cos(\theta)$ establishes the connection 
to the polar angle $\theta$.

This partial differential equation is solved by a separation of variables and the full solution (Ref.
~\cite{Aragon:JCP_82:1985}) is obtained as an expansion in terms of spheroidal wave functions $\Ps_n^m$ of degree $n$
and order $m$:
\begin{align}
  G_\vec{k}(z, \phi, t| z_0, \phi_0) = \sum_{m=-\infty}^\infty \sum_{n = m}^\infty \frac{2n +
  1}{4\pi}\frac{(n - m)!}{(n + m)!} \Ps_n^m(z,\gamma^2)\Ps_n^m(z_0,\gamma^2) e^{\img m (\phi -
  \phi_0)} e^{-\Gamma_n^m t}.
\end{align}
where the real parameter $\gamma^2$ and the characteristic decay constants $\Gamma_n^m \equiv \Gamma_n^m(\gamma^2)$
depend only on the diffusion coefficients $D_\parallel$, $D_\perp$, and $D_\text{rot}$ and magnitude of the wave vector
$k$. The exact relations for both parameters are determined as solutions of the spheroidal wave equation
\begin{align}
  \partial_z[(1-z^2)\partial_z \Ps_n^m] + \bigg[\lambda_n^m(\gamma^2) + \gamma^2(1 - z^2) -
\frac{m^2}{1-z^2}\bigg] \Ps_n^m = 0\ ,
\end{align}
where in our case $\gamma^2 = k^2 \Delta D/D_\text{rot}$ and the spheroidal
eigenvalue $\lambda_n^m \equiv \lambda_n^m(\gamma^2)$ determines the decay rate $\Gamma_n^m = D_\parallel k^2 +
D_\text{rot} \lambda_n^m$~\cite{NIST:DLMF,Olver:2010:NHMF}.
Only contributions of order $m = 0$ remain after averaging over all initial orientations and integration over all final
orientations:
\begin{align}
  F(\vec{k}, t) = \sum_{n = 0}^\infty \frac{2n + 1}{4} \bigg[\int_{-1}^1\diff{z}\ \Ps_n^0(z,\gamma^2)\bigg]^2 e^{-\Gamma_n^0 t} 
\end{align}
Since the spheroidal wave functions display the symmetry property, $\Ps_n^0(-z,\gamma^2) = (-1)^n \Ps_n^0(z,\gamma^2)$, only
functions with even degree contribute to the intermediate scattering function after integration over $z$, we can restrict the sum to even $n$.
Expanding the spheroidal wave functions $\Ps_{n}^0$ in terms of Legendre polynomials $\text{P}_{n}^0$,  
\begin{align}
 \Ps_{n}^0(z, \gamma^2) &= \sum_{k=-\lfloor n/2\rfloor}^\infty (-1)^k a_{n,k}^0(\gamma^2)\text{P}_{n+2k}^0(z)\ ,
\end{align}
and perform the remaining integration via the orthonormality condition of the Legendre polynomials:
\begin{align} \label{eq:coefficient_spheroidal}
 \int_{-1}^1\diff{z}\ \Ps_n^0(z,\gamma^2) =  \int_{-1}^1\diff{z}\ \Ps_n^0(z,\gamma^2) \text{P}_0^0(z)
=  2 (-1)^n a_{n,-n/2}^0(\gamma^2)\ .
\end{align}
Hence, the intermediate scattering function reduces to a sum over
exponentially decaying functions with decay constants $\Gamma_{n}^0$ and amplitudes given by
the first coefficient of the spheroidal wave function $a_{n,-n/2}^0$, the overlap between the zeroth order spheroidal
wave function of degree $n$ and the zeroth degree Legendre polynomial:
\begin{align} \label{eq:intermediate_scattering_spheroidal}
   F(\vec{k}, t) = e^{-k^2 D_\parallel t}\sum_{\substack{n = 0\\ n \text{ even}}}^\infty (2n + 1)
(a_{n,-n/2}^0)^2 e^{-\lambda_n^0 D_\text{rot} t}\ .
\end{align}
Methods for the computation of the eigenvalues $\lambda_n^0$ and the coefficients $a_{n,-n/2}^0$ are readily
available in the literature~\cite{Hodge:JMP_11:1970,Falloon:JPA:2003, NIST:DLMF}. The evaluation of the sum is truncated
when the spheroidal eigenvalue $\lambda_n^0$ becomes positive and the sum is close to unity for time zero.  For the
highest densities and wave numbers, $\gamma^2 \sim 10^6$, more than $10^3$ terms contribute and
multi-precision floating point arithmetic is mandatory~\cite{Falloon:JPA:2003}.  

In the highly entangled regime $D_\text{rot}\to 0$ and thus $\gamma^2\to\infty$, the spheroidal eigenvalue 
approaches $\lambda_n^0 = -\gamma^2 + (2n+1)\gamma + \mathcal{O}(1)$ and the spheroidal wave functions
$\text{Ps}_n^0(z,\gamma^2)$ for fixed $n$ converge in mean square to the eigenfunctions of the harmonic oscillator~\cite{Meixner:Springer:1954}:
\begin{align}
\biggl(\frac{2n+1}{2}\biggr)^{1/2}\text{Ps}_n^0(z,\gamma^2) = \biggl(\frac{\sqrt{\gamma}}{\sqrt{\pi} 2^n
n!}\biggr)^{1/2} \exp\Bigl(-\frac{\gamma z^2}{2}\Bigr) H_n(\sqrt{\gamma} z) + \mathcal{O}(\gamma^{-1})
\end{align}
where $H_n(\cdot)$ denote the Hermite polynomials. By analogy to the harmonic oscillator, the width of the 
eigenfunctions approaches $\sqrt{n/\gamma}$, and the integral in Eq.\eqref{eq:coefficient_spheroidal} can be safely
extended to the real line for $n \ll \gamma$. 
The remaining integral for even $n$ can be evaluated exactly by employing the generating function of the Hermite polynomials
in terms of a generalized binomial coefficient
\begin{align}
(a_{n,-n/2}^0)^2  \simeq \sqrt{\frac{\pi}{\gamma}}\frac{(-1)^{n/2}}{2n+1} \binom{-1/2}{\phantom{-}n/2} .
\end{align}
Thus, by Newton's binomial series, the sum [Eq.~\eqref{eq:intermediate_scattering_spheroidal}] can be evaluated in closed form and the intermediate scattering functions assumes
the approximate form
\begin{align}
F(k,t) &= \sqrt{\frac{\pi}{\gamma}} e^{-k^2 D_\perp t}e^{-D_\text{rot}\gamma t} \sum_{\substack{n = 0\\ n \text{
even}}}^{n \ll \gamma} (-1)^{n/2} \binom{-1/2}{\phantom{-}n/2} e^{-2n D_\text{rot} \gamma t}
\\ \label{eq:intermediate_scattering_approx} & =\sqrt{\frac{\pi}{\gamma}} e^{-k^2 D_\perp t} e^{-D_\text{rot}\gamma t} \Bigl[1-e^{-4 D_\text{rot}\gamma t}\Bigr]^{-1/2}
= e^{-k^2 D_\perp t} \sqrt{\frac{\pi}{\gamma}} \frac{1}{\sqrt{2\sinh(2 D_\text{rot}\gamma t)}}
\end{align}
for times $D_\text{rot}\gamma^2 t \gtrsim 1$, such that only terms $n \ll \gamma$ contribute to the sum. In particular, one infers the terminal relaxation rate 
$k^2 D_\perp + D_\text{rot}\gamma$ of the intermediate scattering function.
Furthermore, for times $D_\text{rot} \gamma t \ll 1$ the result can be simplified further
\begin{align} \label{eq:intermediate_asymptotic_supp}
F(k, t) = \frac{e^{-k^2 D_\perp t}}{\sqrt{4 k^2 \Delta D t/\pi}} ,
\end{align}
Thus in the regime $1 \lesssim D_\text{rot}\gamma^2 t \lesssim \gamma$ the intermediate scattering
function displays a characteristic power-law decay with exponent $-1/2$ in a broad time window.

The origin of the tail can be easily understood as a consequence of a pure sliding motion.
Since the tail becomes manifest for times where the needle barely rotates, $D_\text{rot} t \simeq 1/\gamma$, 
we approximate $D_\text{rot} \approx 0$ in this regime, such that the Smoluchowski equation simplifies to a first order differential equation:
\begin{align}
 \partial_t G_\vec{k} = - k^2 [D_\parallel - \Delta D (1 - z^2)]G_\vec{k} = - k^2[D_\perp + \Delta D z^2] G_\vec{k} .
\end{align}
The solution is directly obtained as $G_\vec{k}(z,\phi,t|z_0,\phi_0) = \exp(-k^2D_\perp t - k^2\Delta D z^2
t)\delta(z,z_0)\delta(\phi,\phi_0)$, describing the motion of a needle where the initial
orientation is preserved for all times.
Then the intermediate scattering function can be evaluated in closed form 
\begin{align}
F(k,t) = \frac{1}{2}e^{-k^2 D_\perp t} \int_{-1}^1\diff z_0\ e^{-k^2 \Delta D z_0^2 t} = e^{-k^2 D_\perp t} \frac{1 -
\text{erfc}(\sqrt{k^2 \Delta D t})}{\sqrt{4 k^2 \Delta D t/\pi}}
\end{align}
where $\text{erfc}(\cdot)$ denotes the complementary error function.
Hence, for large $k^2\Delta D t = \gamma^2 D_\text{rot} t \gtrsim 1$, we recover
Eq.~\eqref{eq:intermediate_asymptotic_supp} .

Let us compare to the solution strategy of Doi and Edwards (ignoring the form factor for the infinitely thin
needle)~\cite{Doi:JCSFT_74:1978}. There, the Smoluchowski-Perrin equation [Eq.~\eqref{eq:smoluchowski-perrin_z_phi}] is
approximated by a time-dependent harmonic oscillator equation after integration over the azimuthal angle which is
correct for polar angles close to the equator. Then, they obtain the intermediate scattering function
\begin{align} \label{eq:intermediate_scattering_doi}
F(k,t) = \frac{1}{2}\int_{-1}^1\diff z\int_{-1}^1\diff z_0\ g_\vec{k}(z,t|z_0),
\end{align}
where the integral extends over all polar angles of the Green function of a harmonic oscillator 
\begin{align}
g_\vec{k}(z,t|z_0) = \frac{\sqrt\gamma \exp(-k^2D_\perp t)}{\sqrt{2\pi\sinh(2\gamma D_\text{rot} t)}}\exp\biggl(-\gamma\frac{\cosh(2\gamma
D_\text{rot} t)(z^2+z_0^2)}{2\sinh(2\gamma D_\text{rot} t)}
+\gamma\frac{z z_0}{\sinh(2\gamma D_\text{rot} t)}\biggr).
\end{align}
Using the generating function of the Hermite polynomials, the Green function can be expressed also as 
\begin{align}
g_\vec{k}(z, t|z_0) = e^{-k^2D_\perp t}\sum_{n=0}^\infty \frac{\sqrt{\gamma}}{\sqrt{\pi}2^n n!}
e^{-\gamma(z^2+z_0^2)/2}H_n(\sqrt{\gamma}z)H_n(\sqrt{\gamma}z_0) e^{-(2n+1)\gamma D_\text{rot} t} ,
\end{align}
which is obtained from the exact solution by replacing the eigenvalue and eigenfunctions by their harmonic oscillator analogue irrespective of the
degree $n$. However, as discussed above, this replacement is only valid for $n\ll \gamma$ and correspondingly, the Green function is accurate only for times
$D_\text{rot}\gamma^2 t \gtrsim 1$ and polar angles close to the equator.

In the original work~\cite{Doi:JCSFT_74:1978}, the integral in Eq.~\eqref{eq:intermediate_scattering_doi} can not be performed in closed form rather an additional
approximation is introduced such that the needle keeps its initial polar angle.
We note that one can formally obtain our result Eq.~\eqref{eq:intermediate_scattering_approx} by extending the integrals in Eq.~\eqref{eq:intermediate_scattering_doi} 
to the real line, which amounts to interchanging limits of the infinite sum with the integrals over the polar angles, irrespective of convergence. 

\vspace*{0.5cm}
\twocolumngrid


\end{document}